\documentclass[11pt]{article}
\usepackage{latexsym,amsmath,units,amssymb}
\usepackage[square,comma,sort&compress,numbers]{natbib}
\usepackage{pstricks}
\renewcommand{\a}{{\alpha}}
\renewcommand{\b}{{\beta}}

\newcommand{\dl}{{\delta}}
\newcommand{\eps}{{\epsilon}}

\newcommand{\m}{{\mu}}
\newcommand{\n}{{\nu}}
\newcommand{\pa}{{\partial}}

\newcommand{\sig}{{\sigma}}
\newcommand{\Sig}{{\Sigma}}
\renewcommand{\r}{{\rho}}
\renewcommand{\th}{{\theta}}
\newcommand{\ze}{{\zeta}}
\newcommand{\la}{{\lambda}}

\newcommand{\om}{{\omega}}
\newcommand{\Om}{{\Omega}}
\newcommand{\ag}{\mathfrak{g}}

\newcommand{\an}{\mathfrak{g}_{\textnormal{\sc nw}}}

\newcommand{\mB}{\textnormal{\sffamily{\textit{B}}}}
\newcommand{\mG}{\textnormal{\sffamily{\textit{G}}}}
\newcommand{\mH}{\textnormal{\sffamily{\textit{H}}}}
\newcommand{\mJ}{\textnormal{\sffamily{\textit{J}}}}

\numberwithin{equation}{section}
\renewcommand{\theequation}{\arabic{section}.\arabic{equation}}

\begin{document}
\thispagestyle{empty}

\begin{center} {\huge\bf  D-branes with Lorentzian signature\\[12pt] 
                       in the  Nappi-Witten model}

\vskip 48pt 
{\Large R. Hern\'andez, G. Horcajada and F. Ruiz Ruiz} 

\vskip 3pt
\emph{Departamento de F\'{\i}sica Te\'orica I, Universidad Complutense de
  Madrid \\ 28040 Madrid, Spain}

\vskip 36pt 
\today
\vskip 18pt
\begin{abstract}
  Lorentzian signature D-branes of all dimensions for the Nappi-Witten string
  are constructed. This is done by rewriting the gluing condition $J_+=FJ_-$
  for the model chiral currents on the brane as a well posed first order
  differential problem and by solving it for Lie algebra isometries $F$ other
  than Lie algebra automorphisms. By construction, these D-branes are not
  twined conjugacy classes. Metrically degenerate D-branes are also obtained.
\end{abstract}
\end{center}

\vspace{60pt} {\sc Keywords:} D-branes, Penrose limit and pp-wave background

\newpage
\vskip 36pt
\section{Introduction}

The approach to understanding D-branes and their properties in terms of the
open strings attached to their worldvolumes has provided a new look on
D-branes.  One of the most remarkable and influential results along this line
is the observation that, for flat spacetime and a globally defined constant
two-form~$\mB$, the \hbox{D-brane} world volume becomes noncommutative upon
quantization~\citep{Chu-Ho-flat,Schomerus,Seiberg-Witten,Seiberg}.  In
particular, if $\mB$ is of magnetic type, space directions do not
commute~\citep{Chu-Ho-flat,Schomerus,Seiberg-Witten}, whereas if $\mB$ is of
electric type, it is the time direction that does not commute with the space
directions~\citep{Seiberg}. Since the three-form~$\mH$ vanishes for
constant~$\mB$, the field equations for the string are the same as for
$\mB=0$. The boundary conditions, however, change, since they involve the
field $\mB$; the annhilation and creation parts of every string mode get
coupled and this coupling leads to noncommutativity upon
quantization. Noncommutativity should then be a general feature for D-branes
in curved backgrounds.

Two important examples of the latter are provided by (i) the family of
\hbox{\emph{pp}-wave} geometries~\cite{Chu-Ho-pp,Horcajada-Ruiz} with also a
globally defined constant $\mB$ that describe the Penrose limits of
\hbox{$\textnormal{AdS}_n\times\textnormal{S}^m$} and
\hbox{$\textnormal{dS}_n\times\textnormal{S}^m$}, and (ii) an
$\textnormal{S}^3$ background with nonzero $\mH$~\citep{ARS-fuzzy,ARS-fluxes}.
In the first case, the three-form $\mH$ vanishes and noncommutativity at the
string endpoints can be established through canonical quantization. In the
second case, canonical quantization is not adequate and noncommutativity is
proved by resorting to the $SU(2)$ WZW formulation of the string background.

More generally, since WZW models are the building blocks of many string
backgrounds, one expects to learn about D-branes and their noncommutative
field theories by looking at open strings on group manifolds. This entails as
a first step the characterization of D-branes in WZW
models~\cite{Klimcik-Severa, Alekseev-Schomerus, Stanciu-3, Stanciu-manifolds,
  Stanciu-note, FS-more,Bachas-Petropoulos, Ribault-Schomerus, Cheung-Freidel,
  HRR}. Such characterization is well understood in some cases. In particular,
it is known that the metrically nondegenerate $R$-twined conjugacy classes of
a WZW group manifold are \hbox{D-branes} for all Lie algebra metric-preserving
automorphisms $R$. These twined conjugacy classes are obtained as the
solutions to a gluing condition $J_+=RJ_-$ that matches the chiral currents
$J_+$ and $J_-$ of the model at the \hbox{D-brane}. Very little is known,
however, if in the gluing condition, instead of a Lie algebra automorphism, an
arbitrary isometry $F$ of the Lie algebra metric is
considered~\cite{HRR}. This is due to the fact that involutivity
(required for the solution to the gluing condition to define a submanifold)
holds trivially for Lie algebra automorphisms, whereas for general isometries
it usually does not. In this latter case, involutivity often requires to
consider isometries $F(g)$ that depend on the group point $g$ at which the
gluing condition must be solved.

The Nappi-Witten model~\citep{Nappi-Witten} is a WZW model describing a
four-dimensional noncompact string background. The twined conjugacy classes
of its group manifold are well understood~\citep{FS-more}.  They provide 
two-dimensional Euclidean D-branes for metric-preserving inner automorphisms
and three-dimensional Lorentzian D-branes for metric-preserving outer
automorphisms. The purpose of this paper is to go beyond and construct
Lorentzian D-branes of dimension one, two, three and four by solving the
gluing condition for isometries other than Lie algebra automorphisms. Our
motivation aims to constructing noncommutative field theories on noncompact
curved backgrounds.

In this paper we apply geometric characterization of D-branes in nonsemisimple
Lie groups along the lines of ref.~\cite{HRR} to the Nappi-Witten model. There
are other ways to approach the study of D-branes in WZW models. In particular,
the so called algebraic program, that uses boundary conformal theory. See
\emph{e.\,g.} ref.~\cite{Schomerus-CQG} and references therein for compact
string backgrounds, ref.~\cite{Schomerus-PhysRep} for noncompact ones and
ref.~\cite{AK} for their use in the Nappi-Witten model.

The paper is organized as follows. In Section 2, we review the semiclassical
characterization of D-branes in a WZW model. The material presented there can
be found elsewhere~\citep{Stanciu-3, Stanciu-manifolds, Stanciu-note}, 
though it emphasizes some points~\citep{HRR} concerning the r\^ole of
Frobenius theorem and involutivity that have gone somewhat unnoticed in the
literature. Section 3 contains a brief account of the Nappi-Witten model,
including a complete characterization of its Lie algebra isometries. Sections
4 and 5 are dedicated to constructing the D-branes of interest. In particular,
worldvolume filling D-branes, metrically degenerate D2-branes and Lorentzian
D1-branes are presented in Section 4, whereas Lorentzian D2 and D0-branes are
exhibited in Section 5. The subject of Section 6 is to recover the boundary
conditions for the string coordinates from the gluing condition for the chiral
currents. Finally, Section 7 collects our conclusions.

\section{Characterization of D-branes in WZW models}

Consider a Lie algebra $\ag$ of dimension ${\tt d}$ over ${\bf R}$ and an
invariant Lie algebra metric $\Om$ defined on it. In a basis $\{T_a\}$, with
commutation relations
\begin{equation}
   [\,T_a,T_b\,] = f_{ab}{}^c\, T_c \,,
\label{CR}
\end{equation}
the metric components $\,\Om_{ab}=\Om(T_a,T_b)$ satisfy
\begin{equation}
   f_{ab}{}^d\,\Om_{dc} = \Om_{ad}\,f_{bc}{}^d \,.
\label{invariance}
\end{equation}
Here we will be interested in isometries of $\Om$. An isometry of $\Om$ is a
linear map $F$ from $\ag$ to $\ag$ such that\,
$\Om(FT_a,FT_b)=\Om(T_a,T_b)$. Writing the action of $F$ on a generator $T_a$
as $\,F(T_a)=T_b\,F^b{}_a$, with $\,F^b{}_a\,$ taking values in ${\bf R}$, the
isometry condition becomes\footnote{In matrix notation, in $F^a{}_b$ the index
  $a$ specifies the row and the index $b$ the column.}
\begin{equation}
  \Om_{ab}  = F^c{}_a\,\Om_{cd}\,F^d{}_b ~~\Leftrightarrow~~
    F^{\rm T}\,\Om\,F=\Om \,.
\label{isometry}
\end{equation}
The isometries of $\Om$ form a subgroup $\textnormal{\sl Iso}(\Om)$ of the
general linear group $\,G\ell\,({\tt d},{\bf R})$. Our conventions for matrix
notation is that the first index, from left to right, labels rows, and the
second one labels columns.

The pair $(\ag,\Om)$ defines a WZW model described by mappings $g$ from the
string worldsheet $\Sig$ to the group manifold $G$ obtained from $\ag$ through
exponentiation. If $G$ is locally parameterized by the string coordinates
$X^\m(\tau,\sig)$, the left-invariant $e^a{}_\m$ and right-invariant
$\bar{e}{\,^a}{}_\m$ vielbeins at $g(X)$ are
\begin{equation}
    g^{-1}\,dg =T_a\, e^a{}_\m\,dX^\m\,,\qquad 
    dg\,g^{-1} = T_a \,\bar{e}{\,^a}{}_\m\,dX^\m\,.
\label{viel}
\end{equation} 
The adjoint action of the group $G$ on the algebra $\ag$ is 
\begin{equation*}
   {\rm Ad}_g (T_a) =  g\,T_a\,g^{-1}= 
       T_b\;\bar{e}^{\,b}{}_\m\,(e^{-1})^{\,\m}{}_a\> 
            ~~~~\Leftrightarrow~~~~  {\rm Ad}_g= \bar{e}\,e^{-1}\,,
\end{equation*}
where $\,{(e^{-1})^\m}{}_a\,$ is the inverse of $\,e^a{}_\m$, defined by
$\,{(e^{-1})^\m}{}_a\,e^b{}_\m=\dl_a{}^b$. The spacetime metric $\mG_{\!\m\n}$
and the three-form $\mH_{\m\n\la}$ specifying the string background are given in
terms of $\Om$ by
\begin{eqnarray}
   & \mG_{\m\n} = \Om\,\big(g^{-1}\pa_\m g\,,\,g^{-1}\pa_\n g\big) & 
    \label{st-metric} \\[3pt]
   & \mH_{\m\n\la} = \Om\, \big(\big[g^{-1} \pa_\m g\,,
                \,g^{-1} \pa_\n g\big]\,,\,g^{-1} \pa_\la g\big)\,. &
\label{st-H}
\end{eqnarray}
In world sheet coordinates \hbox{$\,\sig^\pm \!= \tau\pm \sig$}, the chiral
currents of the model read
\begin{equation*}
        \mJ_-(\sig^-)=g^{-1}\pa_-g\,, \qquad \mJ_+(\sig^+)=-\,\pa_+g\,g^{-1}
\end{equation*}
and satisfy $\pa_+\mJ_-=\pa_-\mJ_+=0$.

A Dp-brane is a $(p+1)$-dimensional submanifold $N$ of $G$ on which an open
string may end. Points in $N$ can be parameterized by the string endpoints
coordinates \hbox{$x^\m(\tau)=X^\m(\tau,\sig)\big\vert_{\pa\Sig}$}, so we will
write $g(x)$. The D-brane can be
specified~\citep{Klimcik-Severa,Stanciu-manifolds,Stanciu-note,HRR} by
\begin{itemize}
\vspace{-\itemsep}\vspace{-\partopsep}
\item[(i)] An isometry $F$ of $\Om$, that in general may depend on $g$,
    and a condition
\begin{equation}
    J_+ = F(g)\, J_- ~~~\textnormal{at}~~~\pa\Sig\,.
\label{GC}
\end{equation}
This condition must define $i=1,\ldots,p+1$ integrable vector fields 
$k_i(x)=k^\m{\!}_i(x)\pa_\m$ that characterize the tangent bundle of the 
submanifold $N$. The fields $k_i(x)$ must define a basis of $T_gN$ for all 
$g(x)$ in $N$. In what follows we will denote by $\a^i$ the local 
coordinates along the directions defined by $k_i$, that is,\, 
$k_i=\pa/\pa\a^i$.  \vspace{-\itemsep}\vspace{-\parsep}
\item[(ii)] A two-form $\omega$ defined on $N$, with components
  $\om_{ij}=\om(k_i,k_j)$ satisfying the following two requirements.  Firstly,
  eq.~(\ref{GC}) must reproduce the usual boundary conditions of the sigma
  model formulation, which in the presence of a D-brane read~\citep{HRR,
    Klimcik-Severa}
\begin{equation}
       \big(k^\m{\!}_i \,\mG_{\m\n}\,\pa_\sig X^\n
           - \om_{ij}\,\pa_\tau \a^j\big)\, \Big\vert_{\pa\Sig} =0
        \qquad i=1,\ldots,p+1\,.
\label{BC}
\end{equation} 
And secondly, $d\omega=\mH\big\vert_N$. Note that the variations of the 
D-brane coordinates $\a^i$ and the string endpoints coordinates $x^\m$ with 
$\tau$ are related by $\pa_\tau\a^i\,k^\m{\!}_i=\pa_\tau\/x^\m$.
\vspace{-\itemsep}\vspace{-\partopsep}
\end{itemize}
We remark that, from the viewpoint of the sigma model, the boundary
conditions take the form~(\ref{BC}).  These are the equations that must be
recovered from eq.~(\ref{GC}).  To avoid confusion, eq.~(\ref{GC}) is called
gluing condition.  

Writing the chiral currents as\, \hbox{$J_{-\!} = T_a\,
  e^{\,a}{\!}_\m\,\pa_-X^\m$} \,and\, \hbox{$J_{+\!} = \!- T_a\,
  \bar{e}^{\,a}\!{}_\m\,\pa_+X^\m$}, multiplying from the left with the
right-invariant inverse vielbein, and using world sheet coordinates $\tau$ and
$\sig$, condition~(\ref{GC}) is written as
\begin{equation}
  \big({\cal F} - 1\big)\,\pa_\tau X \big\vert_{\pa\Sig}
  = \big({\cal F} + 1\big)\, \pa_\sig X \big\vert_{\pa\Sig} \,.
\label{GC-ts}
\end{equation}
Here ${\cal F}$ stands for
\begin{equation}
  {\cal F}^{\,\m}{}_\n 
       = -\,(\bar{e}^{\,-1})^{\,\m}{}_a\, F^{\,a}{}_b\;e^{\,b}{}_\n
       ~~~\Leftrightarrow~~~ {\cal F}(x)= -\, \bar{e}^{-1} F(g)\,e
\label{calF}
\end{equation}
and is called matrix of boundary conditions. ${\cal F}$ is only defined at
$\pa\Sig$ and depends on $x^\m$ through the vielbeins\, $e(x)$ \,and\,
$\bar{e}^{\,-1}(x)$ \,and the isometry $F\big(g(x)\big)$. 

For any $g$ in $G$, the only motions compatible with conditions~(\ref{GC-ts})
are along the curves tangent to the vector
fields~\citep{Stanciu-manifolds,Stanciu-note,HRR}
\begin{equation*}
    t_U(g)= FUg-gU\,,\quad U\in\ag\,.
\end{equation*}
Since\, $U=U^aT_a$ \,for all $U$ in $\ag$ and $\{T_a\}$ is a basis, it is
enough to consider the fields
\begin{equation*}
   t_a(g)= FT_ag-gT_a\,.
\end{equation*}
At every $g$, the fields $t_a(g)$ define a space of tangent directions
\begin{equation*}
  \Pi_{g} = \textnormal{\sl Span }\{\,{t}_a(g)\,\} 
\end{equation*}
contained in the tangent space $T_gG$ at $g$ to the whole manifold $G$ . If
the tangent planes $\Pi_g$ have dimension\, $p+1$ \,for all~$g$
in~$G$, their collection defines a $(p+1)$-dimensional distribution on $G$,
\begin{equation}
  \Pi  = \{\,\Pi_g\!:\,\textnormal{dim}\,\Pi_g=p+1,\;g\in\/G\,\}\,.
\label{distribution}
\end{equation}
Multiplication of $t_a(g)$ from the left with $g^{-1}$ \,gives\, \hbox{$g^{-1}t_a =
  \textnormal{Ad}_{g^{-1}}FT_a - T_a$}. The planes $\Pi_g$ have dimension
$p+1$ for all $g$ if and only if\; \hbox{$\textnormal{Ad}_{g^{-1}}F -1$} \;has
rank $p+1$ for all $g$. It may occur that $\Pi_g$ does not have constant
dimension over $G$, so that $\Pi$ is not a distribution on $G$. In this
case, since the open string endpoints flow along the integral curves of $t_a$,
it is enough to have a distribution
\begin{equation*}
   \Pi\,'=\{\,\Pi_g\!:\,\textnormal{dim}\,\Pi_g=p+1,\;g\in\/G\,'\,\}
\end{equation*}
on a submanifold $G\,'$ of $G$, provided $G\,'$ contains all such curves. See
ref.~\citep{HRR} for details and Sections 4 and 5 for examples.

For $\Pi_g$ to be at all $g$ in $G^{\,\prime}\!$, not just a tangent plane,
but the tangent space to a submanifold $N_{p+1}$ of $G^{\,\prime}\!$, the
distribution $\Pi^{\,\prime}$ must be integrable. According to Frobenius
theorem, $\Pi^{\prime}$ is integrable if and only if the vector fields $t_a$
are involutive. That is, if and only if the commutator $\,[\,t_a,t_b]\,$ of
any two fields $t_a$ and $t_b$ taking values in $\Pi^{\,\prime}$ also takes
values in $\Pi^{\prime}$. This amounts to the existence of functions\,
$c_{ab}{\!}^c(g)$ \,such that
\begin{equation}
    [\,t_a(g),t_b(g)\,] = c_{ab}{\!}^c(g)\;t_c(g)
\label{invol}
\end{equation}
for all $g$ in $G^{\,\prime}$. The distribution $\Pi^{\prime}$ is the tangent
bundle of $N_{p+1}$.

The field $t_a(g)$ is the sum of a right-invariant vector field $\,X_{\!\rm
  R}g$, with $X_{\!\rm R}=FT_a$, and a left-invariant one $gX_{\!\rm L}$, with
$\,X_{\!\rm L}=\!-T_a$. Its action on a differentiable function~$f$ defined on
$G$ is easily computed from the actions of $\, X_{\!\rm R}g\,$ and
$\,gX_{\!\rm L}$, given by
\begin{equation*}
  X_{\!\rm R}g\, \big(f(g)\big)  
       = \frac{d}{dt}~f\big( e^{tX_{\!\rm R}} g\,\big)\bigg\vert_{t=0}~,~\qquad
  gX_{\!\rm L}\, \big(f(g)\big) 
        = \frac{d}{dt}~f\big( g\,e^{tX_{\!\rm L}}\, \big)\bigg\vert_{t=0}\,.
\end{equation*}
If $g(x)$ is parameterized by coordinates $x^\m$, the components of
$\,X_{\!\rm R\,}g\,$ and $\,gX_{\!\rm L}\,$ are
\begin{align}
  X_{\!\rm R\,}g & = X_{\!\rm R}^a\,T_a\, g = X_{\!\rm R}^a\,
     {({\bar{e\,}}^{-1})^\m}_a\,\pa_\m 
   \label{right-field} \\[3pt]
  gX_{\!\rm L}\! & = X_{\!\rm L}^a\,gT_a 
    = X_{\!\rm L}^a\,{(e^{-1})^\m}_a\,\pa_\m \,.
    \label{left-field}
\end{align}
This gives 
\begin{equation}
    t_a(x) = FT_ag-gT_a = \big[\,{(\bar{e\,}^{-1})^\m}_b F^b{}_a  
                   - \,{(e^{-1})^\m}_a\,\big]\,\pa_\m 
           = {t}^{\,\m}{\!}_a(x)\,\pa_\m.
\label{vect-fields}
\end{equation}
The rank of the matrix\, \hbox{${t}^{\,\m}{\!}_a(x)$} \,at $g(x)$ is the
dimension of $\Pi_g$. Note that the $a$-th column of this matrix is formed by
the components of the tangent vector ${t}_a(x)$.

Let us restrict our attention for the time being to isometries $F$ that are
constant over $G$. In this case, equation~(\ref{invol}) takes the simpler
form~\citep{HRR}
\begin{equation}
    -\,\big[FT_a,FT_b\big]\,g +  g\,\big[T_a,T_b\big] 
         = c_{ab}{\!}^c(g)\>(FT_cg - gT_c)\,.
\label{involutive-constant}
\end{equation}
A solution to this equation is provided by $\,F=R^{\,\Om}\,$ and\,
\hbox{$c_{ab}{}^c(g)=-f_{ab}{}^c$}, with $R^{\,\Om}$ a Lie algebra automorphism
satisfying eq.~(\ref{isometry}) and $f_{ab}{}^c$ the Lie algebra structure
constants. This is trivial since, by definition, Lie algebra automorphisms~$R$
satisfy
\begin{equation}
      R\,[T_a,T_b] = [RT_a,RT_b] \,.
\label{auto}
\end{equation}
The restriction to automorphisms $R^{\,\Om}$ complying with
eq.~(\ref{isometry}) comes from the observation that for a general Lie algebra
not all automorphisms $R$ are isometries. Automorphisms
fulfilling~(\ref{isometry}) condition are called $\Om$-preserving. The vector
fields\, \hbox{$t_a\!=R^\Om T_ag -gT_a$} \,are very easy to integrate and give
for the submanifold $N$ the $R^\Om$-twined conjugacy classes of the
group~$G$~\citep{Alekseev-Schomerus,Stanciu-manifolds,Stanciu-note},
\begin{equation*}
  N={\cal C}(R,g_0) = \big\{ e^{\,R^\Om\/V}g_0\,e^{-V}\!:\,V\in\ag \big\}
\end{equation*}
where $g_0$ is an arbitrary group element that accounts for the integration
constants. 

There are suggestions~\citep{Stanciu-3,Stanciu-manifolds,Stanciu-Tseytlin}
that $F=\!-R^\Om$, with $R^\Om$ an $\Om$-preserving constant automorphism, may
solve involutivity and, hence, may lead to D-branes. For semisimple Lie
algebras, however, it has been proved~\citep{HRR} that this is not the
case. In the following sections we examine this problem for the Nappi-Witten
model, a typical example of nonsemisimple WZW model. We find that
$\,F=-\!R^{\,\Om}$ does not define D-branes if $R^{\,\Om}$ is constant, but it
does if $R^{\,\Om}$ is conveniently taken to depend on $g$. The choice of the
$g$-dependence of $R^{\,\Om}(g)$ is indicated by the requirement of the
constancy of the rank of the matrix\, \hbox{${t}^{\,\m}{\!}_a(x)$} In Sections
4 to 6 a several examaples are presented.

\section{The Nappi-Witten model: a brief review}

The Nappi-Witten model~\citep{Nappi-Witten} is constructed upon a nonsemisimple
Lie algebra $\an$ whose exponentiation gives a group manifold
$G_{\textnormal{\sc nw}}$ describing a nontrivial four-dimensional string
background of \emph{pp}-type. The algebra has dimension four and generators
$\{P_1,P_2,J,T\}$ with commutation relations
\begin{equation}
   [J,P_M]=\eps_{MN} P_N, \qquad [P_M,P_N] = \eps_{MN} T,
      \qquad [T,P_i]=[T,J]=0,  \qquad M,N=1,2\,.
\label{NW-algebra}
\end{equation}
It is the central extension of the Eucliedean algebra in two dimensions, $T$
being the central charge. We will use the labeling
\begin{equation*}
  T_1=P_1, \qquad T_2=P_2, \qquad T_3=J, \qquad T_4=T \,,
\end{equation*}
so that 
\begin{equation}
 f_{12}{}^4 = f_{31}{}^2= f_{23}{}^1 = 1 \,.
\label{fabc}
\end{equation}
The most general invariant metric $\Om$  can be found by solving
eqs.~(\ref{invariance}). It reads~\citep{Nappi-Witten}
\begin{equation}
   \Om=k \begin{pmatrix} 1 & 0 & 0 & 0 \\
                             0 & 1 & 0 & 0 \\
                             0 & 0 & b & 1 \\
                             0 & 0 & 1 & 0 \end{pmatrix} ~,
\label{NW-Omega}
\end{equation}
where $k$ and $b$ are arbitrary real parameters. The parameter $k$ can be
absorbed in the coupling constant in front of the classical WZW action, so
that it can be set equal to one without loss of generality. As concerns $b$,
it can be set to zero by the following redefinition of the Lie algebra
generators:
\begin{equation*}
 P^{\,\prime}_M= P_M, \quad J^{\,\prime}= J -\frac{b}{2}\>T\,,\quad T^{\,\prime}=T\,.
\end{equation*}
Indeed, under such transformations, the Lie algebra commutators~(\ref{NW-algebra})
remain unchanged and the metric $\Om$ takes the form in~(\ref{NW-Omega}) with
$b=0$. We thus set $k=1$ and $b=0$ without loss of generality.

\vspace{9pt}
\noindent\underline{\sl Lie algebra isometries}

\vspace{6pt}
The isometries $F$ of $\Om$ are the solutions to equation~(\ref{isometry}). To
find them, it is most convenient to write $\Om$ in eq.~(\ref{NW-Omega}) as
$\,\Om =  M^{\rm T}\,\eta\,M$, with $M$ the matrix
\begin{equation*}
  M = \begin{pmatrix} 1 & 0 & 0 & 0\\[2pt]
                        0 & 1 & 0 & 0 \\[2pt]
   0 ~&~ 0 ~&~ \dfrac{\sqrt{2}}{2} & \dfrac{\sqrt{2}}{2}\\[9pt]
   0 & 0 & \dfrac{\sqrt{2}}{2} & -\dfrac{\sqrt{2}}{2} 
     \end{pmatrix} 
\end{equation*}
and $\,\eta=\textnormal{\sl diag}\,(+,+,+,-)$. Eq.~(\ref{isometry}) then becomes
\begin{equation*}
   {\big(MFM^{-1}\big)}^{\rm T}\,\eta\,\big(MF\,M^{-1}\big) = \eta\,.
\end{equation*}
This is solved by\, $MFM^{-1}$ \,an arbitrary element of $O(3,1)$, so the
isometry group is
\begin{equation*}
   \textnormal{\sl Iso}(\Om) = \big\{ M^{-1}\Lambda\,M\!: ~ 
    \Lambda \in O(3,1)\,\big\} \,.
\end{equation*}
Note that $M$ is not an isometry.

\vspace{9pt}
\noindent\underline{\sl Lie algebra automorphisms}

\vspace{6pt} The automorphisms of the Nappi-Witten algebra can be found by
solving eqs.~(\ref{auto}). Using the structure constants $\,f_{ab}{}^c\,$ in
eq.~(\ref{fabc}), it is straightforward to see that eqs.~(\ref{auto}) only
have two solutions, $R_+$ and $R_-$, given by
\begin{equation}
  R_\pm (\r_0,\r,\phi,\th)=  \begin{pmatrix}
    \r_0\cos\phi & \mp \r_0\sin\phi 
       & \mp\,\frac{\displaystyle \r}{\displaystyle \r_0} \cos\th & 0 \\[3pt]
    \r_0\sin\phi &  \pm \r_0\cos\phi 
       & -\frac{\displaystyle \r}{\displaystyle \r_0} \sin\th & 0  \\[3pt]
    0 & 0 & \pm 1 & 0 \\[3pt]
    \r\cos(\th\mp \phi)  & \r \sin(\th\mp\phi) & \ze  & \pm \r_0^2 
          \end{pmatrix} \,.
\label{gen-aut}
\end{equation} 
The parameters $\r_0,\,\r,\,\phi,\,\th$ and $\ze$ can take any values on the
ranges
\begin{equation*}
   \r_0 > 0\,,\qquad  \r\geq 0 \,,\qquad  0\leq \phi\,,\,\th < 2\pi \,,
   \qquad -\infty<\ze<\infty\,.
\end{equation*}
As is well known, the automorphisms $R_-$ and $R_+$ form a group,
$\textnormal{Aut}(\an)$. There are a few observations concerning
automorphisms and isometries that we find relevant:

$\bullet$ $\textnormal{Aut}(\an)$ is not $O(3,1)$ nor a subgroup of it. This
would require $R_\pm$ to satisfy\, $R_\pm^{\rm T} \,\eta\, R_\pm\!=\eta$,
but this only occurs for $\,\r_0\!=1$, $\,\ze=\r=0$.

$\bullet$ Not every isometry is a Lie algebra automorphism. Take for example
\begin{equation}
   F_0= M^{-1}\eta M= \begin{pmatrix}  1 & 0 & 0 & 0 \\[3pt]
                                      0 & 1 & 0 & 0 \\[3pt]
                                      0 & 0 & 0 & 1 \\[3pt] 
                                      0 & 0 & 1 & 0 \end{pmatrix}\,.
\label{counter-example}
\end{equation}
By construction, $F_0$ is an isometry but does not have the
form~(\ref{gen-aut}), hence is not an automorphism\footnote{As a matrix
  $F_0$ is equal to $\Om$, but they have different index structures:
  $(F_0){}^a{}_b$ and $\Om_{ab}$.}.

$\bullet$ Conversely, not every automorphism is an isometry. For this to be
the case, $R_\pm$ must satisfy\, $R_\pm^{\rm T}\,\Om\,R_\pm=\Om$. Some algebra
shows that this is so if and only if $\,\r_0{\!}=1\,$ and $\,\ze=\mp\r^2/2$. From
now on, we denote by $R^\Om_\pm$ automorphisms of this type,
\begin{equation}
  R^\Om_\pm(\r,\phi,\th) =  \begin{pmatrix}
    \cos\phi &  \mp \sin\phi & \mp \r \cos\th  & 0 \\[3pt]
    \sin\phi & \pm \cos\phi &  - \r \sin\th  & 0 \\[3pt]
     0 & 0 & \pm 1 & 0 \\[3pt]
    \r\cos(\th\mp \phi) &  \r \sin(\th\mp\phi) 
          & \mp\,{\displaystyle\frac{\r^2}{2}} & \pm 1 \end{pmatrix} \,.
\label{iso-aut}
\end{equation}
They form the subgroup $\textnormal{Aut}_\Om(\an)$ of $\Om$-preserving
automorphisms. 

\newpage
\vspace{9pt}
\noindent\underline{\sl The spacetime group manifold}

\vspace{6pt}
A group element $g$ can be parameterized as~\citep{Nappi-Witten}
\begin{equation*}
   g(x_M,u,v) = e^{ x_M P_M} \,e^{uJ}\, e^{vT}\,,  \qquad M=1,2
\end{equation*}
in terms of real coordinates $x_M,u,v$. In this parameterization, the identity
element is $e=g(0,0,0)$, while the group law takes the form
$\,g(x)\,g(x')=g(x'')$, with
\begin{align*}
    x''_M & = x_M\!+\cos u\, x'_M\! - \sin u\, \eps_{MK}\,x'_K\\
    u''&=u+u'\\
    v''& =  ~v+v' + \tfrac{1}{2}\,\cos u\,\eps_{MK}\, x_M \,x'_K\!  
         + \tfrac{1}{2}\,\sin u\, x_M\, x_M'\,.
\end{align*}
The inverse $g^{-1}$ of $g$ reads
\begin{equation*}
   g(x_M,u,v)^{-1} 
       = g\,(-\cos u~x_M -\sin u~\eps_{MK}\,x_K,\>-u,\>-v\,)\,.
\end{equation*}
The left and right-invariant vielbeins follow easily from their
definition~(\ref{viel}). They read 
\begin{equation*}
 e^a{}_\m =\begin{pmatrix} \cos u &  \sin u & 0 & 0 \\[3pt]
                          -\sin u & \cos u & 0 & 0 \\[3pt]
                           0 & 0 & 1 & 0 \\[3pt]
       {\displaystyle \frac{x_2}{2}} & -{\displaystyle \frac{x_1}{2}} 
                           & 0 & 1  \end{pmatrix}
  \,,~\qquad
  \bar{e}^{\,a}{}_\m =\begin{pmatrix} 
                   1 & 0     & x_2  & 0   \\[3pt]
                   0 & 1     & -x_1 & 0   \\[3pt]
                   0 & 0     & 1    & 0   \\[3pt]
  - {\displaystyle \frac{x_2}{2}} & {\displaystyle \frac{x_1}{2}} 
                   & -\, {\displaystyle \frac{1}{2}}\> (x_1^2+x_2^2) & 1
                    \end{pmatrix}  \,.
\end{equation*}
Some simple algebra gives then for the adjoint action of the group on the Lie
algebra
\begin{equation}
      \textnormal{Ad}_{g}=\bar{e}\, e^{-1}  = \begin{pmatrix}
       \cos u &  -\sin u & x_2  & 0 \\[3pt]
       \sin u & \cos u & - x_1& 0 \\[3pt]
       0 & 0 & 1 & 0 \\[3pt]
       x_1\sin\/u - x_2 \cos\/u &  x_1 \cos\/u + x_2\sin\/u  
           & -\,{\displaystyle \frac{1}{2}}\> (x_1^2+x_2^2)  & 1 
                                             \end{pmatrix} \,.
\label{Adg}
\end{equation}
It follows that $\,R^\Om_+\,U =\textnormal{Ad}_{h}U$ for all $U$ in
the Nappi-Witten algebra, with $h$ a group element with coordinates
\begin{equation*}
  x_{1\!}=\r\sin\th\,, \qquad x_{2\!}=-\r\cos\th\,, \qquad  
   u=\phi\,, \qquad v~\textnormal{arbitrary} \,. 
\end{equation*}
This implies that automorphisms of type $R^\Om_+$ are inner. As
regards automorphisms of type $R_-^\Om$, they are outer since there is no
group element $h$ such that $\,R^\Om_-U=\textnormal{Ad}_{h}U\,$ for all $U$.

The spacetime metric and the WZW three-form are given by
eqs.~(\ref{st-metric}) and~(\ref{st-H}). In the coordinates that we are using
they take the form
\begin{align}
   ds^2 & = dx_1^2 + dx_2^2 
      + (x_2\,dx_1 - x_1\,dx_2)\,du + 2\,du\,dv \label{G-NW}\\[3pt]
   \mH & = dx_1 \wedge dx_2 \wedge du\,. \label{H-NW}
\end{align}

\section{Filling D-branes and D-strings from general isometries}

We are interested in finding if isometries of the form $F=\!-R^{\,\Om}$ define
D-branes. We will consider both constant and $g$-dependent automorphisms. It
is convenient to separately discuss inner and outer automorphisms.

\subsection{\bf Case $\boldsymbol{F=\!-R^\Om_{\textnormal{\bf inner}}}$}

Using eqs.~(\ref{iso-aut}) and ~(\ref{vect-fields}),  for $\,F=\!-R^\Om_+$, we
obtain the following vector fields~$t_a$:
\begin{align}
  t_1(x) &= -\,(\cos\phi +\cos u)\,\pa_1 - (\sin \phi + \sin u)\,\pa_2 
     \nonumber \\
      & +\frac{1}{2}\> \big[\, x_1\, (\sin\phi - \sin u) 
        - x_2\, (\cos\phi -\cos u) 
        - 2\r \cos(\th-\phi) \,\big]\,\pa_v  \label{t1} \\[6pt]
  t_2(x) &=  (\sin \phi + \sin u)\,\pa_1 - (\cos\phi + \cos u)\,\pa_2 
        \nonumber \\
      & + \frac{1}{2}\> \big[\,  x_1\, (\cos\phi - \cos u) 
            +  x_2\, (\sin\phi -\sin u) 
            - 2\r \sin(\th-\phi) \,\big]\,\pa_v \label{t2} \\[6pt]
  t_3(x) &= ~  (x_2 + \r\cos\th)\,\pa_1 - (x_1-\r\sin\th)\, \pa_2 
   - 2\,\pa_u
   - \frac{\r}{2}\>(x_1\sin\th -x_2\cos\th -\r)\,\pa_v  \label{t3}\\[6pt]
  t_4(x) & =  -\,2\,\pa_v\,. \label{t4}
\end{align}
They involve the four derivatives $\pa_1,\,\pa_2,\,\pa_u$ and $\pa_v$. In
particular, $\,\pa_u$ only enters $t_3$ with constant coefficient, so the
motion defined by~$t_3$ covers the whole range for $u$.  A simple calculation
shows that
\begin{equation} 
   \textnormal{det}\,\big( {t}^{\,\m}{\!}_a\big) 
     = 8\,\big[\, 1 + \cos(\phi-u)\, \big]\,.
\label{det-inner}
\end{equation}

\vspace{9pt}
\noindent
\underline{\sl Nonexistence of D-branes for constant $F=-\!R^{\,\Om}_+$}

\vspace{6pt} Let us first consider that $R^{\,\Om}_+$ does not depend on
$g(x)$, so the parameters $\r,\,\phi,\,\th$ are constant. For points $g(x)$
with $\,u\neq\phi+(2n+1)\pi$, the determinant~(\ref{det-inner}) does not
vanish and the tangent planes
\begin{equation*}
   \Pi_{g(x)}=\textnormal{\sl  Span}\,\{ {t}_1,\, {t}_2,\, 
         {t}_3,\,{t}_4\}  ~~\textnormal{for}~~  u\neq\phi+(2n+1)\pi
\end{equation*}
have dimension four. At points $g(x)$ with\, $u=\phi+(2n+1)\pi$, however, the
determinant~(\ref{det-inner}) vanishes. In a neighborhood of these points the
fields $t_1$ and $t_2$ become
\begin{align*}
   t_1(x) & = \big[ \,x_1\,\sin\phi - x_2\,\cos\phi - 
           \r\,\cos(\th-\phi)\,\big]\,\pa_v \\
   t_2(x) & = \big[ \,x_1\,\cos\phi + x_2\,\sin\phi - 
           \r\,\sin(\th-\phi)\,\big]\,\pa_v \,,
\end{align*}
while $t_3$ and $t_4$ remain as in~(\ref{t3}) and~(\ref{t4}). The fields
$t_1,\,t_2$ and $t_4$ define then the same tangent direction, namely
$\pa_v$, so the tangent planes are spanned by $t_3$ and $t_4$,
\begin{equation*}
   \Pi_{g(x)}=\textnormal{\sl  Span}\,\{{t}_3,\,{t}_4\}
      ~~\textnormal{for}~~ u=\phi+(2n+1)\pi\,,
\end{equation*}
and have dimension $2$. Hence, the dimension of $\Pi_{g(x)}$ is not the same
for all $g(x)$ in $G_{\textnormal{\sc nw}}$, the collection of tangent planes
$\Pi_{g(x)}$ is not a distribution on $G_{\textnormal{\sc nw}}$ and Frobenius
theorem does not apply. The same conclusion can be reached by studying
  the rank of\; \hbox{$\textnormal{Ad}_{g^{-1}}F -1$} \,(see the Appendix).

One may consider the submanifold 
\begin{equation*}
    G^{\,\prime}_{\textnormal{\sc nw}}=G_{\textnormal{\sc nw}}-\{g(x)\!:\, 
          u=\phi+(2n+1)\pi\}
\end{equation*}
that results from removing from $G_{\textnormal{\sc nw}}$ the closed set of
group elements $g(x)$ with $u=\phi+(2n+1)\pi$. The collection
\begin{equation*}
  \Pi\,'= \big\{\Pi_{g(x)}\!:\; g(x) \in\/  G^{\,\prime}_{\textnormal{\sc nw}}\big\}
\end{equation*}
is now a distribution of dimension four on $G^{\,\prime}_{\textnormal{\sc
    nw}}$. Furthermore, having maximal dimension, it is trivially involutive.
The manifold $G^{\,\prime}_{\textnormal{\sc nw}}$ cannot, however, be accepted
as a D-brane.  The reason is that it does not contain the integral curves of
$t_3$, which connects points $g(x')$ with\, \hbox{$u'\neq\phi+(2n+1)\pi$} with points
$g(x)$ with\, \hbox{$u=\phi+(2n+1)\pi$} that are not in
$G^{\,\prime}_{\textnormal{\sc nw}}$, thus contradicting the idea that the
string endpoints lie on the D-brane. The gluing
condition~(\ref{GC}) does not define then a D-brane for constant
$\,F=\!-R^{\,\Om}_+$.

\vspace{9pt}
\noindent\underline{\sl Filling D-branes and D-strings for nonconstant
  $F=-\!R^{\,\Om}_+$} 

\vspace{6pt} The situation is very different if $R^{\,\Om}_+$ depends on
$g(x)$.  Assume that we take 
\begin{equation}
   F_4=R^{\,\Om}_+(\r,\phi,\th)\,, \qquad \phi(u)=u+\phi_0\,, \qquad
     \phi_0=\textnormal{\sl const}\neq\/(2n+1)\pi\,.
\label{filling}
\end{equation}
The matrix\, $t^{\m}{\!}_a$ \,has now nonvanishing determinant for all $g(x)$,
so the collection $\Pi$ of all the tangent planes $\Pi_{g(x)}$ is a
distribution of dimension four on $G_{\textnormal{\sc nw}}$. Having maximal
dimension, $\Pi$ is trivially involutive and is thus the tangent bundle of
$G_{\textnormal{\sc nw}}$ itself. In Section 6 we show that the gluing
condition~(\ref{GC-ts}) for $F_4$ in eq.~(\ref{filling}) with $\r$ and $\th$
constant can be written as a boundary condition~(\ref{BC}) for a two-form
$\omega$ such that $d\omega=\mH$ on $G_{\textnormal{\sc nw}}$. The gluing
condition for such an $F_4$ hence defines a filling D-brane.

Consider now the isometry
\begin{equation}
   F_2=R^{\,\Om}_+(\r,\phi,\th)\,, \qquad \phi(u)=u-\pi\,.
\label{D-string}
\end{equation}
The determinant\, \hbox{$\textnormal{det}(t^{\m}{\!}_a)$}\,then vanishes for
all $g(x)$. In the neighborhood of any $g(x)$, the fields $t_1$ and $t_2$ read
\begin{align*}
    t_1(x) & = -\,\big[\,x_1\sin\/u 
             - x_2\cos\/u - \r\cos(\th-u) \,\big]\,\pa_v \\
    t_2(x) & = -\,\big[\,x_1\cos\/u 
             + x_2\sin\/u - \r\sin(\th-u) \,\big]\,\pa_v \,,
\end{align*}
while $t_3$ and $t_4$ remain as in~(\ref{t3}) and~(\ref{t4}). The only partial
derivative that occurs in $t_1,\,t_2$ and $t_4$ is $\pa_v$, so they define the
same tangent direction. The tangent planes \hbox{$\Pi_{g(x)}$} have dimension
two for all $g(x)$ and are spanned by $t_3$ and $t_4$. Their collection
$\Pi_2$ is hence a distribution of dimension two on $G_{\textnormal{\sc nw}}$
and Frobenius theorem can be used. It is trivial that $[t_3,t_4]=0$, so the
distribution is integrable. $\Pi_2$ defines a family of two-dimensional
submanifolds $N_2$ whose tangent space at all $g(x)$ is
$\,T_{g(x)}N_2=\Pi_{g(x)}$. In Section 6, we show that the gluing condition
can be recast as a boundary condition for a two-form $\omega$ defined on
$N_2$.  Such form trivially satisfies $d\omega=\mH\big\vert_{N_2}$, so the
submanifolds $N_2$ are \hbox{D1-branes} and provide a foliation of
$G_{\textnormal{\sc nw}}$.

Redefining $v\to v-bu/2$ and using eq.~(\ref{G-NW}), we have that
\begin{equation*}
   \mG(t_3,t_3) = -\,\big[ \big(x_1-\r\sin\th\big)^2  
     + \big(x_2+\r\cos\th\big)^2\,\big]< 0\,,   \qquad 
   \mG(t_3,t_4)=8>0\,, \qquad    \mG(t_4,t_4)=0\,.
\end{equation*}
Every submanifold $N_2$ in the family has then Lorentzian signature and is a
D-string.  If $\a^1$ and $\a^2$ parameterize the integral curves of 
$k_1(x)=t_3(x)$ and $k_2(x)=t_4(x)$ in eqs.~(\ref{t3}), the D-string is 
formed by points\, $x^\m(\a^1,\a^2)$ \,such that
\begin{equation}
   dx^\m = k^{\m}{}_1(x)\,d\a^1 + k^{\m}{}_2(x)\,d\a^2\,.
\label{diff-eqs}
\end{equation}
The induced metric on the D-string takes the form
\begin{equation}
   ds_2^2 = \mG(k_1,k_1) \, {(d\a^1)}^2 + \mG(k_1,k_2)\,d\a^1\,d\a^2\,. 
\label{newD1}
\end{equation}
Assume now that $\r$ and $\th$ depend on $x_1,\,x_2$ and $u$, but not on
$v$. Noting that eqs.~(\ref{diff-eqs}) imply that $x_1,\,x_2$ and $u$ only
depend on $\a^1$, we conclude that $\mG(k_1,k_1)$ only depends on $\a^1$ and 
thus eq.~(\ref{newD1}) is a \emph{pp}-wave metric in 1+1 dimensions.

To find the metric coefficient $\mG(k_1,k_1)$ as a function of $\a^1$, some further assumptions on $\r$ and $\th$ are necessary. For example, for
$\r$ and $\th$ constant, integrating eqs.~(\ref{diff-eqs}), we obtain
\begin{align*}
   x_1 &= \r\sin\th +r_0\cos(\a^1+\varphi_0) \\ 
   x_2 &=-\r\cos\th -r_0\sin(\a^1+\varphi_0) \\ 
   u &= -2\a+u_0 \\
   v &= -2\b + \frac{r_0\r}{2}~\cos(\a^1+\varphi_0+\th) + v_0\,,
\end{align*}
with $r_0,\,\a_0,\,u_0$ and $v_0$ integration constants.  The D-string metric
is then\, \hbox{ $-r_0^2\, {(d\a^1)}^2 + 8\,d\a^1\,d\a^2$}. To the best of our
knowledge, the family~(\ref{newD1}) of D-strings has gone unnoticed in the
literature.

\subsection{\bf Case $\boldsymbol{F=\!-R^\Om_{\textnormal{\bf outer}}}$}

For $\,F=\!-R^\Om_-$, there are only three nonzero vector fields $\,t_a$,
given by
\begin{align}
  t_1(x) & = -\,(\cos\phi +\cos u)\,\pa_1 - (\sin \phi + \sin u)\,\pa_2 
     \nonumber \\
      & +\frac{1}{2}\> \big[\, x_1\, (\sin\phi - \sin u) 
        - x_2\, (\cos\phi -\cos u) 
        - 2\r \cos(\th+\phi) \,\big]\,\pa_v  \label{tou1} \\[4.5pt]
  t_2(x) &= -\, (\sin \phi - \sin u)\,\pa_1 + (\cos\phi - \cos u)\,\pa_2 
        \nonumber \\
      & - \frac{1}{2}\> \big[\,  x_1\, (\cos\phi + \cos u) 
            +  x_2\, (\sin\phi  +\sin u) 
            + 2\r \sin(\th+\phi) \,\big]\,\pa_v \label{tou2} \\[4.5pt]
  t_3(x) &= - (x_2 + \r\cos\th)\,\pa_1 + (x_1+\r\sin\th)\, \pa_2 
   - \frac{\r}{2}\>(x_1\sin\th  + x_2\cos\th +\r)\,\pa_v \,. \label{tou3}
\end{align}
They involve $\pa_1,\,\pa_2$ and $\pa_v$, but not $\pa_u$, hence they define
motions that leave $u$ constant. The matrix $\,t^{\m}{\!}_a\,$ of the
coefficients is now $\,3{\scriptstyle\times}3$, with $\,a=1,2,3\,$ and
$\,\m=1,2,v$.  A straightforward calculation gives
\begin{equation}
  \textnormal{det}\,\big(t^{\m}{\!}_a \big) = -2\, \big[k(x_1,x_2)\big]^{2}\,,
\label{ksquared}
\end{equation}
where $\,k(x_1,x_2)\,$ is the function of $x_1$ and $x_2$
\begin{equation}
  k(x_1,x_2) = \big(x_1+\r\sin\th\big)\,\cos\!\Big(\frac{\phi+u}{2}\Big)
       + \big(x_2+\r\cos\th\big)\,\sin\!\Big(\frac{\phi+u}{2}\Big)\, .
\label{det-outer}
\end{equation}

\vspace{9pt}
\noindent
\underline{\sl Nonexistence of D-branes for constant $F=\!-R_-^{\,\Om}$}

\vspace{6pt}
At points $g(x)$ with\, $k(x_1,x_2)\neq\/0$, the determinant~(\ref{ksquared})
does not vanish and the fields~$t_a$ define three-dimensional tangent planes
$\Pi_{g(x)}$. By contrast, for $g(x)$ with $\,k(x_1,x_2)=0$, the
determinant~(\ref{ksquared}) vanishes. It is straightforward to see that the
rank of the matrix $\,t^{\m}{\!}_a$ is one in this case, so the corresponding
tangent planes $\Pi_{g(x)}$ have dimension one.  The collection of all the
planes $\Pi_{g(x)}$ is not a distribution on $G_{\textnormal{\sc nw}}$ and
Frobenius theorem cannot be used.  This conclusion can also be reached by
studying the rank of\, $\textnormal{Ad}_{g^{-1}}F-1$ \,for $F=\!-R^{\,\Om}_-$
(see the Appendix).

One could think of removing from $G_{\textnormal{\sc nw}}$ the locus of points
for which $k(x_1,x_2)=0$. The resulting submanifold
$G^{\,\prime}_{\textnormal{\sc nw}}$ then does not include all the points
accessible to the string endpoints, since $k(x_1,x_2)=0$ can be reached from
$k(x_1,x_2)\neq\/0$ through the motions defined by the fields $t_a$. Hence the
gluing condition~(\ref{GC}) does not define a D-brane for constant
$\,F=\!-R^{\,\Om}_-$.

\vspace{9pt}
\noindent\underline{\sl D2 and D0-branes for nonconstant $F=\!-R_-^{\,\Om}$}

\vspace{6pt} Let us take now $\r,\,\phi$ and $\th$ in
$R^{\,\Om}_-(\r,\phi,\th)$ functions of $x_1$ and $x_2$ such that\,
\hbox{$k(x_1,x_2)=k_0$}, with $k_0$ a nonzero constant. The fields $t_1,\,t_2$
and $t_3$ define then three-dimensional tangent planes $\Pi_{g(x)}$ for all
$g(x)$ in $G_{\textnormal{\sc nw}}$, whose collection $\Pi_3$ is a
three-dimensional distribution on $G_{\textnormal{\sc nw}}.$ Alternatively,
$\Pi_3$ is a three-dimensional distribution on any constant $u=u_0$
three-plane\, $N_{u_0}=\{g(x)\in G_{\textnormal{\sc nw}}\!:\, u=u_0 \}$. The
distribution $\Pi_3$ is trivially involutive and defines the tangent bundle of
the three-plane $u=u_0$. This plane has two spacelike directions and one
lightlike direction, but no timelike direction, so the metric is degenerate.
In Section 6, the gluing condition is written as the boundary condition for a
two-form $\omega$ defined on the three-plane $u=u_0$ such that
$d\omega=H\big\vert_{u_0}\!=0$, thus proving that the planes $u=u_0$ are
degenerate D2-branes.

We next consider $\r,\,\phi$ and $\th$ functions of $x_1$ and $x_2$ such that
$k(x_1,x_2)=0$ for all $x_1$ and $x_2$. In the neighborhood of any point
$g(x)$ in $G_{\textnormal{\sc nw}}$ the vector fields $t_a$
in~(\ref{tou1})-(\ref{tou3}) take the form
\begin{align}
   t_1(x) &= -\,(\cos\phi +\cos\/u)\,
             \big( \pa_1 + \dfrac{\r}{2}\,\cos\th\,\pa_v\big)
             -(\sin\phi + \sin\/u)
                \big( \pa_2 - \frac{\r}{2}\,\sin\th\,\pa_v\big) 
             \label{D0-1} \\[3pt]
   t_2(x) &= -\,(\sin\phi - \sin\/u)\,
             \big( \pa_1 + \dfrac{\r}{2}\,\cos\th\,\pa_v\big)
             + (\cos\phi -\cos\/u)\,
             \big( \pa_2 - \frac{\r}{2}\,\sin\th\,\pa_v\big) 
             \label{D0-2}\\[3pt] 
   t_3(x) &= -\,(x_2 + \r\cos\th)\,
             \big( \pa_1 + \dfrac{\r}{2}\,\cos\th\,\pa_v\big) 
            + ( x_1 + \r\sin\th)\,
            \big( \pa_2 - \frac{\r}{2}\,\sin\th\,\pa_v\big)\,.
            \label{D0-3}
\end{align}
It is very easy to convince oneself that these vectors define a
one-dimensional distribution $\Pi_1$ on~$G_{\textnormal{\sc nw}}$. Being
one-dimensional, $\Pi_1$ is trivially involutive. Its integral curves 
$N_1$ are spacelike since
\begin{equation*}
   \mG(t_a,t_a) > 0  ~~~ \textnormal{for}~~~ t_a\neq 0,\quad a=1,2,3\,. 
\end{equation*}
To give their explicit form, some further assumptions on the dependence of
$\r,\,\phi$ and $\th$ on $x_1,\,x_2$ and $u$ must be made. Let us give some
examples.

Take\,  $x_{1\!}+\r\sin\th=0$ \,and\, $\phi+u=\phi_0\neq\/2n\pi$. Condition\,
$k(x_1,x_2)=0$ \,implies\,$x_{2\!}+\r\cos\th=0$. These three equations
define $\r$ and $\th$ in terms of $x_1$ and $x_2$, and $\phi$ in terms of
$u$. The fields $t_1,\,t_2$ and $t_3$ become
\begin{equation*}
  t_1 =\!-\,2\,\cos\Big(u-\frac{\phi_0}{2}\Big)\,t_0\,, \qquad~
  t_2 = 2\,\sin\Big(u-\frac{\phi_0}{2}\Big)\,t_0 \,,\qquad~
  t_3=0\,,
\end{equation*}
where $t_0$ stands for 
\begin{equation*}
  t_0 = \cos\big(\frac{\phi_0}{2}\big)\,
           \big( \pa_1 + \dfrac{\r}{2}\,\cos\th\,\pa_v\big)
      +\, \sin\big(\frac{\phi_0}{2}\big)\,
           \big( \pa_2 - \frac{\r}{2}\,\sin\th\,\pa_v\big) \,.
\end{equation*}
It is clear that $t_1$ and $t_2$ do not vanish simultaneously and specify the
same direction at every $x^\m$. The integral curves are in this case
$v=x_1^0\,x_2-x_2^0\,x_1 + v_0$, with $x_1^0,\,x_2^0$ and $v_0$ integration
constants.

Assume now that\, $x_{1\!}+\r\sin\th$ \,and\, $x_{2\!}+\r\cos\th$ \,do not
vanish simultaneously. The field $t_3$ is then nonvanishing and the integral
curves are formed by $x^\m(\a)$, with $u=u_0$ \,and\, $x_1,\,x_2$ \,and $v$
the solutions to
\begin{equation*}
   \dfrac{dx_1}{d\a}= -\,(x_2 + \r\cos\th)\,, ~\qquad
   \dfrac{dx_2}{d\a}= x_1 + \r\sin\th\,, ~\qquad
   \dfrac{dv}{d\a}= - \dfrac{\r}{2}\> 
           \big(x_1\sin\th + x_2\cos\th + \r\big) \,,
\end{equation*}
where $\a$ is a parameter along the curve. For\, $\r=0$, the integral curves are
circles\, \hbox{$x_1^2+x_2^2=r_0^2$} \,of arbitrary radius $r_0$ located
on any two-plane $(u\!=\!u_0\,, v\!=\!v_0)$. A simple solution for $\r\neq 0$
is provided by $\r=-x_1/\sin\th_0$, with $\th=\th_0\neq\/n\pi$ constant
and $\phi=\!-u_0$. In this case, the integral curves are
parabolas $v=-\frac{1}{4}\cot\th_0\,x_1^2+v_0$ on any two-plane $(x_2\!=x_2^0\,,
u\!=\!u_0)$. 

In any case, being one-dimensional,\, $d\omega=\mH\big\vert_{N_1}$ \,is
trivially satisfied, and the curves $N_1$ are D0-branes.

\section{D2 and D0-branes with Lorentzian signature}

In the previous Section we have constructed D3 and D1-branes with Lorentzian
signature by integrating the gluing condition for some $g$-dependent
isometries $F=\!-R^{\,\Om}$. Here we construct D2-branes and D0-branes, also
with Lorentzian signature, for $g$-dependent isometries $F\neq\pm\/R^{\,\Om}$.

Since the product of two isometries is an isometry and $F_0$ in
eq.~(\ref{counter-example}) is an isometry, $F=F_0R^{\,\Om\!}$, with
$R^{\,\Om}$ an arbitrary metric-preserving automorphism, is also an
isometry. Isometries of this type do not have the form
$\pm\/R^{\,\Om}$. Let us take for $R^{\,\Om}$ an inner automorphism, so that
we will be considering\, $F=F_0R_+^{\,\Om}$. The corresponding vector
fields~$t_a$ are
\begin{align*}
t_1 - \r \cos(\phi-\th)\; t_4 & =  (\cos\phi-\cos\/u) \,\pa_1 
    + (\sin\phi-\sin\/u) \,\pa_2 \\
    & + \frac{1}{2} \,\big[\,2\r\,\cos(\phi-\th) - x_1\,(\sin\phi+\sin\/u)
        + x_2\,(\cos\phi+\cos\/u)\,\big]\,\pa_v \\
t_2 + \r \sin(\phi-\th)\; t_4& = -\, (\sin\phi-\sin\/u) \,\pa_1 
    + (\cos\phi-\cos\/u) \,\pa_2 \\
    &~~~\,- \frac{1}{2} \,\big[\,2\r\,\sin(\phi-\th) + x_1\,(\cos\phi+\cos\/u)
    + x_2\,(\sin\phi+\sin\/u)\,\big]\,\pa_v \\
t_3  + \frac{1}{2}\>(\r^2 +2)\, t_4& =  
   -(x_2+\r\cos\th)\,\pa_1 + (x_1-\r\sin\th)\,\pa_2 
   +\frac{\r}{2}\,(x_1\sin\th-x_2\cos\th-\r\,)\,\pa_v \\
t_4 & = - x_2\,\pa_1+x_1\,\pa_2 +\pa_u - \pa_v 
\end{align*} 
From these expressions it follows that\,
$\textnormal{det}\,(t^{\,\m}{\!}_a)=0$. This indicates that there are no
filling D-branes for the isometry that we are considering. We look for
D-branes of lower dimension.

Let us take\, $\phi(u)=u$ \,and\, $\r=0$. Since the parameter $\th$ always
occurs in $R_+^{\,\Om}$ through\, $\r\sin\th$ \,and\, $\r\cos\th$, see
eq.~(\ref{iso-aut}), we can set without loss of generality $\th=0$. The
isometry $F$ then reads
\begin{equation}
     F_3(u) = F_0\,R_+^{\,\Om}(0,u,0)\,,
\label{F3}
\end{equation}
and the fields $t_a$ become 
\begin{align}
    t_1 & = ( - x_1 \sin\/u + x_2 \cos\/u)\,\pa_v \label{sim1}\\[3pt]
    t_2 & = (-x_1\cos\/u - x_2\sin\/u)\,\pa_v \label{sim2}\\[3pt]
    t_3 & = - \pa_u + \pa_v \label{sim3}\\[-1pt]
    t_4 & = - x_2\,\pa_1+x_1\,\pa_2 + \pa_u - \pa_v\,. 
    \label{sim4}
\end{align}
The rank of the matrix\, $t^{\m}{\!}_a$ \,of coefficients is now three for\,
\hbox{$x_1^2+x_2^2\neq\/0$}, and one for\, $x_{1\!}=x_{2\!}=0$. We discuss
these two instances separately.

\vspace{9pt}
\noindent \underline{\sl D2-branes}

\vspace{6pt}
Consider the four-dimensional submanifold
\begin{equation*}
   G_4= \big\{\, g(x)\in G_{\textnormal{\sc nw}}\!:~ 
           x_1^2+x_2^2\neq\/0\,\big\}\,.
\end{equation*}
The group elements $g(x)$ that are not in $G_4$ have $x_{1\!}=x_{2\!}=0$. As
both $x_1$ and $x_2$ approach zero, the coefficients of $\pa_1$ and $\pa_2$ in
eqs.~(\ref{sim1})-(\ref{sim4}) vanish, so the fields $t_a$ do not connect 
points in $G_4$ with points outside $G_4$. In other words, the integral 
curves of $t_a$ stay in $G_4$. Furthermore, since the rank of the matrix 
$t^{\m}{\!}_a$ is three for all $g$ in $G_4$, the fields $t_a$ define a 
three-dimensional distribution $\Pi_3$ on $G_4$ formed by the tangent 
planes $\Pi_{g}=\textnormal{\sl Span}\,\{t_2,t_3,t_4\}$. We may alternatively 
take
\begin{equation}
  \Pi_{g(x)}= \textnormal{\sl Span}\,\big\{\, 
     k_1:=-x_2\pa_1+x_1\pa_2\,,~\, k_2:=\pa_u\,, ~\,k_3:=\pa_v \big\} \,.
\label{k123}
\end{equation}
The commutator of any two fields\, $k_1,\,k_2,\,k_3$ \,vanishes, thus implying
that they are involutive. According to Frobenius theorem, $\Pi_3$ is the
tangent bundle of a family of three-dimensional submanifolds $N_3$ foliating
$G_4$.  If $\a^1,\a^2$ and $\a^3$ parameterize the integral curves of\,
$k_1,\,k_2$ and $k_3$, a manifold $N_3$ in the family is formed by points\,
$x^\m(\a^1,\a^2,\a^3)$ \,such that
\begin{equation*}
   dx^\m=k^{\m}{\!}_1\,d\a^1 +k^{\m}{\!}_2\,d\a^2  + k^{\m}{\!}_3\,d\a^3\,.
\end{equation*}  
Integrating these equations we obtain
\begin{equation}
    N_3\!:\quad 
    \begin{array}{l}
     x_1=r_0\cos(\a^1+\varphi_0) \\
     x_2=r_0\sin(\a^1+\varphi_0) 
    \end{array}\,, \quad u= \a^2 +u_0\,, \quad v=\a^3 +v_0\,,
\label{N3}
\end{equation}
with $r_0\!>0,\,\varphi_0,\,u_0$ and $v_0$ arbitrary integration constants. Note that\,
$r_{0\!}=0$ \,corresponds to\, $x_{1\!}=x_{2\!}=0$, which is excluded from 
$G_4$ and will be discussed below. The induced metric on $N_3$ is 
\begin{equation}
  ds_3^2 = r_0^2\,d\a^1\,(d\a^1-d\a^2) + 2\,d\a^2\,d\a^3\,.
\label{3pp}
\end{equation} 
For every $r_0^2\!>\!0$, this is a \emph{pp}-wave in $2+1$ dimensions. In
Section 6 it is shown that the gluing condition for the isometry $F_3$ can be
cast as a boundary condition with an admissible two-form $\omega$ defined on
$N_3$, thus ensuring that $N_3$ is a D2-brane.

\vspace{9pt}
\noindent \underline{\sl D0-branes}

\vspace{6pt}
Let us now consider the two-dimensional submanifold
\begin{equation*}
  G_2= \big\{\,g(x)\!\in\!\/G_{\textnormal{\sc nw}}\!:~ x_{1\!}=x_{2\!}=0\,\big\}\,.
\end{equation*}
For $g$ in $G_2$, the fields $t_1$ and $t_2$ in~(\ref{sim1}) and (\ref{sim2})
vanish, while $t_3$ and $t_4$ in~(\ref{sim3}) and (\ref{sim4}) are
proportional to each other and define a one-dimensional distribution $\Pi_1$
on~$G_2$.  Having dimension one, $\Pi_1$ is trivially involutive. The integral
curves of $t_4$ have\, $x_{1\!}=x_{2\!}=0$ \,and $u$ and $v$ such that
\begin{equation*}
   \dfrac{du}{d\a} = 1\,, \quad \dfrac{dv}{d\a} = - 1\,, 
\end{equation*}
with $\a$ a parameter along the curve. Integration gives $v+u =c_0$, with
$c_0$ an arbitrary integration constant. These curves are timelike since\,
$\mG(t_4,t_4)\!=\!-2<0$. Furthermore, the induced metric on them is\,
$ds_{1\!}^2=\!-2\,d\a^2$. In Section 6, we prove that the gluing condition for
$F_3$ with\, $ x_{1\!}=x_{2\!}=0$ \,can be written as a boundary condition
with $\omega=0$, hence trivially satisfying\, $d\omega=\mH$ \,on
\,$x_1\!=x_2\!=0$. These timelike lines are then D0-branes.

\section{Comparison with the sigma-model approach}

In Sections~4 and~5 we have integrated the gluing condition for a variety of
isometries. We have anticipated that, in every one of the case considered, the
resulting submanifold $N$ was a D-brane since the corresponding gluing
condition could be written as a sigma model boundary condition with a two-form
$\omega$ defined on $N$ such that $d\omega=\mH\big\vert_N$. Let us show this
here.

We first note that there always exists a two-form $\omega$ defined on
$N$ such that any gluing condition can be written as a boundary condition. For
all $g$ in $N$, $\om$ is specified~\citep{HRR} by its action on tangent
vectors $t_a=FT_ag-gT_a$ in $T_gN$ as
\begin{equation}
    \omega\big( FT_ag-gT_a\,,\,FT_bg-gT_b\big) 
          = \mG\big( FT_ag-gT_a\,,\,FT_bg+gT_b\big)\,,
\label{B}
\end{equation}
where we note the sign change in the second argument on the right hand side.
Recall that the linearly independent vector fields $k_i$~$(i=1,\ldots,p+1)$
that span $T_gN$ are linear combinations of $t_a$ and that the components of
$\om$ are $\om_{ij}=\omega(k_i,k_j)$, so that the two form $\om$ is completely
determined by eq.~(\ref{B}). A separate issue is whether $\om$ satisfies
$d\omega=\mH\big\vert_N$. For D-branes of dimension one and two,
$d\omega=\mH\big\vert_N=0$ is trivial. Dimension three and larger must be
discussed case by case. We concentrate on these cases.

\vspace{9pt}
\noindent\underline{\sl Filling D-brane} 

\vspace{6pt} We start with the isometry $F_4$ in eq.~(\ref{filling}). The
submanifold $N_4$ obtained by integrating the gluing condition was the whole
group $G_{\textnormal{\sc nw}}$.  Computation of the corresponding matrix
${\cal F}_4$ in eq.~(\ref{calF}) and substitution in eq.~(\ref{GC-ts}) gives
after some algebra
\begin{align}
   \pa_\sig X_1\,\big\vert_{\pa\Sig} & = 
     \!- \tan\big(\frac{\phi_0}{2}\big)\;\pa_\tau\/x_2 
     + \frac{1}{2}\,
       \big[\, \big(x_1-\r\sin\th\/\big)\,\tan\big(\frac{\phi_0}{2}\big)
          - x_2 - \r\cos\th \,\big] \,\pa_\tau\/u \label{GC1} \\[4pt]
   \pa_\sig X_2\,\big\vert_{\pa\Sig} & =
       \tan\big(\frac{\phi_0}{2}\big)\;\pa_\tau\/x_1 
     + \frac{1}{2}\, \big[\, x_1-\r\sin\th 
            +\big(x_2+\r\cos\th\/\big)\,\tan\big(\frac{\phi_0}{2}\big)\,\big]
                    \,\pa_\tau\/u \label{GC2} \\[4pt]
   \pa_\sig U\,\big\vert_{\pa\Sig} & = 0 \label{GC3} \\
   2\,\pa_\sig V\,\big\vert_{\pa\Sig} & =  \big[x_2+\r\cos\th
        +\r\sin\th\tan\big(\frac{\phi_0}{2}\big) \big] \,\pa_\tau\/x_1
     -  \big[\,x_1-\r\sin\th
     +\r\cos\th\tan\big(\frac{\phi_0}{2}\big)\, \big] \,\pa_\tau\/x_2 \nonumber\\
    & + \frac{1}{2}\,\big[\,x_1^2 +x_2^2 -\r\,(x_1\sin\th-x_2\cos\th)
        + \r\,\tan\big(\frac{\phi_0}{2}\big)\,
          (x_1\cos\th+x_2\sin\th)\, \big] \,\pa_\tau\/u\,. \label{GC4}
\end{align} 
These are the gluing conditions for the chiral currents written in terms of
$\pa_\tau\/x^\m$ and $\pa_\sig\/X^\m\big\vert_{\pa\Sig}$. We want to compare
them with the sigma model boundary conditions~(\ref{BC}).

Since\, $N_{4\!}=G_{\textnormal{\sc nw}}$ and\, $T_gG_{\textnormal{\sc nw}}$
\,is spanned at all $g$ by the four vector fields $k_i=\dl^\m{\!}_i\pa_\m$,
the boundary conditions~(\ref{BC}) can be written as
\begin{equation}
     \big( \mG_{\!\m\n}\, \pa_\sig\/X^\n\ -
     \omega_{\!\m\n}\,\pa_\tau\/x^\n\big)\big\vert_{\pa\Sig} = 0\,. 
\label{BC-filling}
\end{equation}
Using the expression for the metric $\mG_{\!\m\n}$ in eq.~(\ref{G-NW}) and
noting that $\pa_\tau\/x^\m$ are arbitrary, it is a matter of simple algebra
to check that the gluing conditions~(\ref{GC1})-(\ref{GC4}) take the form of
the boundary conditions~(\ref{BC-filling}) for any two-form $\omega$
with
\begin{align} 
   \omega_{v\m}&=0\\
   \omega_{12} & = \!- \tan\big(\frac{\phi_0}{2}\big) \label{B-filling-1}\\
   \omega_{1u} & =  \frac{1}{2}\,
       \Big[ \big(x_1-\r\sin\th\/\big)\,\tan\big(\frac{\phi_0}{2}\big)
          - x_2 - \r\cos\th \,\Big] \label{B-filling-2}\\[4pt]
   \omega_{2u} & =\frac{1}{2}\, \Big[ x_1-\r\sin\th 
            +\big(x_2+\r\cos\th\/\big)\,\tan\big(\frac{\phi_0}{2}\big)\Big] \,.
   \label{B-filling-3}
\end{align} 
It is a question of algebra to check that these equations can as well be
obtained by using~(\ref{B}). So far no restriction has been placed on $\r$ and
$\th$ in $F_4$. By taking them such that
\hbox{$d\omega=dx_1\wedge\/dx_2\wedge\/du=\mH$}, we conclude that the gluing
condition for $F_4$ defines a filling D-brane.  The simplest way to accomplish
this is to choose $\r$ and $\th$ constant.

It is known~\citep{HRR} that different isometries may define the same
submanifold $N$ but not all of them admit a two-form $\omega$ on $N$ such that
$d\omega=\mH\big\vert_N$. Let us illustrate this with the filling D-brane at
hand. We start by recalling~\citep{HRR} that, given an isometry $F$, it is
always possible to define a new isometry
\begin{equation}
   F^{\,\prime}=\textnormal{Ad}_g\,F^{\,-1}\,\textnormal{Ad}_g\,.
\label{new-F}
\end{equation}
The gluing condition for $F^{\,\prime}$ is integrable if and only if it is for
$F$, in which case they both yield the same submanifold $N$. The gluing
  condition~(\ref{GC-ts}) for  $F^{\,\prime}$ reads
\begin{equation*}
   \big({\cal F^{\,\prime}} - 1\big)\,\pa_\tau X \big\vert_{\pa\Sig}
   = \big({\cal F^{\,\prime}} + 1\big)\, \pa_\sig X \big\vert_{\pa\Sig} \,.
\end{equation*}
Noting that ${\cal F}^{\,\prime}={\cal F}^{-1}$ and multiplying from the left
with ${\cal F}$, it becomes
\begin{equation*}
  \big({\cal F} - 1\big)\,\pa_\tau X \big\vert_{\pa\Sig}
   = - \big({\cal F} + 1\big)\, \pa_\sig X \big\vert_{\pa\Sig} \,.
\end{equation*}
This is the same condition as for $F$, except for a negative sign in front of
the partial derivatives $\pa_\sig\/X\big\vert_{\pa\Sig}$. We now take $F_4$
and consider the corresponding $F_4^{\,\prime}$. The gluing conditions for
$F_4^{\,\prime}$ are then as in eqs.~(\ref{GC1})-(\ref{GC4}) with a negative
sign in front of every $\pa_\sig\/X\big\vert_{\pa\Sig}$. To recover the sigma
model boundary condition~(\ref{BC-filling}), we must take
$\omega^{\,\prime}\!=\!-\omega$, with $\omega$ as in
eqs.~(\ref{B-filling-1})-(\ref{B-filling-3}). This in turn implies that\,
$d\omega^{\,\prime}\!=\!-\mH$.  We conclude that the gluing condition for
$F_4^{\,\prime}$, though integrable, does not define a D-brane.

\vspace{9pt}
\noindent\underline{\sl Degenerate D2-branes}

\vspace{6pt} In Subsection 4.2, the planes $u=u_0$ were obtained upon
integration of the gluing condition for an isometry
$F=\!-R_-^{\,\Om}(\r,\phi,\th)$ with parameters $\r,\,\phi$ and $\th$ such
that\, $k(x_1,x_2)$ \,in eq.~(\ref{det-outer}) took a constant value
$k_{0\!}\neq\/0$ for all $x_1$ and $x_2$. For simplicity we set $\r=0$. The
condition\, $k(x_1,x_2)=k_0$ \,then reads
\begin{equation}
     x_1\cos\Big(\frac{\phi+u_0}{2}\Big) 
        + x_2 \sin\Big(\frac{\phi+u_0}{2}\Big) =k _0
\label{deg-k}
\end{equation}
and the isometry\, $F=\!-R_-^{\,\Om}(0,\phi,0)$ \,becomes a function of $x_1$
and $x_2$. Other choices for $\r$ are treated similarly.

Calculation of the corresponding ${\cal F}_3$ and substitution in
eq.~(\ref{GC-ts}) provides the following gluing conditions:
\begin{align}
    0 & = \pa_\tau\/u  \label{deg-u} \\[2pt]
    \pa_\sig\/X_1 +\frac{x_2}{2}~\pa_\sig\/ U\Big\vert_{\pa\Sig} & = 
        \frac{2}{k_0}\>\sin\Big(\frac{\phi+u_0}{2}\Big)\,\pa_\tau\/v 
        \label{deg-1}\\[3pt]
    -\,\pa_\sig\/X_2 + \frac{x_1}{2}~\pa_\sig\/ U\Big\vert_{\pa\Sig} & = 
        \frac{2}{k_0}\>\cos\Big(\frac{\phi+u_0}{2}\Big)\,\pa_\tau\/v 
        \label{deg2-}\\[3pt]
   -\, \pa_\sig\/U\Big\vert_{\pa\Sig} & = \frac{2}{k_0}\>
      \Big[ \sin\Big(\frac{\phi+u_0}{2}\Big)\,\pa_\tau\/x_1
        - \cos\Big(\frac{\phi+u_0}{2}\Big)\,\pa_\tau\/x_2\Big]  \,. 
        \label{deg-3}
\end{align}
It is very easy to check that eqs.~(\ref{deg-1})-(\ref{deg-3}) can be written
as the\, $i\!=\!1,2,3$ \,boundary conditions that result from taking
$k_1=\pa_1,\,k_2=\pa_2$ and $k_3=\pa_v$ in eqs.~(\ref{BC}) for 
$\om$ given by
\begin{equation}
    \om_{12}=0\,, \qquad  
    \om_{13}=\frac{2}{k_0}\>\sin\Big(\frac{\phi+u_0}{2}\Big) \,,\qquad 
    \om_{23}= -\, \frac{2}{k_0}\>\cos\Big(\frac{\phi+u_0}{2}\Big)\,.
\label{B-degenerate}
\end{equation}
This expression for $\om$ can also be obtained by taking $F=\!-
R_-^{\,\Om}(0,\phi,0)$ in eq.~(\ref{B}). Eqs.~(\ref{deg-k})
and~(\ref{B-degenerate}) imply that $d\om=0$, hence\,
$d\om=\mH\big\vert_{u_0}\!=0$.

\vspace{9pt}
\noindent\underline{\sl Lorentzian D2 branes}

\vspace{6pt} We close by considering the isometry $F_3(u)$ in
eq.~(\ref{F3}). In Section 5 we distinguished two cases: $x_1^2+x_2^2
=r_0^2\neq\/0$ \,and\, $x_{1\!}=x_{2\!}=0$.  In the first one, integration of
the gluing condition resulted in the three-dimensional \emph{pp}-wave in
eqs.~(\ref{N3}) and~(\ref{3pp}), whose tangent space is spanned by the vector
fields $k_1,\,k_2$ and $k_3$ in eq.~(\ref{k123}).  It is straightforward to
show, either by direct computation or by using eq.~(\ref{B}), that the gluing
condition for ${\cal F}_3$ can be recast as boundary conditions with a
two-form $\om$ given in components, by
\begin{equation}
    \om_{12} = \om(k_1,k_2) = \dfrac{r_0^2}{2} - 2  
     \,, \qquad
    \om_{13} = \om(k_1,k_3) = - 2\,, \qquad
    \om_{23} = \om(k_2,k_3) =  1\,.
\label{BN3}
\end{equation}
It is clear that $d\om=0$. On the other hand, since\, $x_1^2+x_2^2=r_0^2$ \,is
a nonzero constant, $dx_1$ and $dx_2$ are not independent and the three-form
$\mH$ vanishes on $N_3$. Hence $d\om= \mH\big\vert_{N_3}$ is trivially
satisfied.

\vspace{9pt}
\noindent\underline{\sl D-strings}

\vspace{6pt} The two-form $\om$ for any 1-dimensional D-brane is trivially
zero. Let us for completeness compute $\om$ for the two-dimensional
\emph{pp}-wave $N_2$ in~(\ref{newD1}) obtained from the isometry $F_2$
in~(\ref{D-string}). Taking $k_1=t_3$ and $k_2=t_4$ in eqs.~(\ref{t3}) and
(\ref{t4}) and using eq.~(\ref{B}), it is straightforward that
$\om_{12}=\om(k_1,k_2)=0$. 

\section{Outlook}  In this paper we have found Lorentzian signature
D-branes of all dimensions for the Nappi-Witten string background. We have
achieved this by formulating the usual gluing condition $J_+=FJ_-$ for the
corresponding WZW chiral currents $J_+$ and $J_-$ and by finding solutions
for Lie algebra isometries $F$ that are not automorphisms, thus generalizing
existing results. Our analysis shows that the methods used to obtain
D-branes for Lie algebra automorphisms work very neatly for more general
cases but require a careful formulation of integrability/involutivity.  In
particular, the occurrence of metrically degenerate D-branes and
coordinate-dependent isometries $F(g)$ are solvable issues.  We envisage
various problems lying ahead. The most inmediate one is perhaps the study of
the low-energy limit of the corresponding effective D-brane actions,
somewhat in the way it is performed in ref.~\citep{ARS-fuzzy}. By doing so,
we expect to learn about noncommutative field theory on curved Lorentzian
D-branes and non-critical strings~\citep{Seiberg}. This may also provide a
way to approach noncommutative solitons as bound states of
strings~\citep{Witten-bound}.  
\section*{Appendix} 
\renewcommand{\theequation}{A.\arabic{equation}}

This appendix contains an alternative derivation to that given in the main
text that the isometries $F=-R_\pm^{\,\Om}$ considered in Subsections 4.1
and 4.2 do not define distributions for constant\,~$R_\pm^{\,\Om}$. 

Since the adjoint group action $\,\textnormal{Ad}_{g^{-1}}$ defines for any
group element $g$ an inner metric-preserving Lie algebra automorphism,
$\,\textnormal{Ad}_{g^{-1}\!}R_\pm^{\,\Om}\,$ is the product of two
automorphisms, hence an automorphism of the same type as $R_\pm^{\,\Om}$. In
fact, eqs.~(\ref{iso-aut}) and (\ref{Adg}) imply that
\begin{equation}
   \textnormal{Ad}_{g^{-1}}\,R_\pm^{\,\Om}(\r,\phi,\th) 
      = {R}_\pm^{\,\Om}(\tilde{\r},\tilde{\phi},\tilde{\th}) \,,
\label{-inner}
\end{equation}
where the parameters $\,\tilde{\r},\,\tilde{\phi},\,\tilde{\th}\,$ depend
on $\r,\,\phi,\,\th$ and $g(x)$ through
\begin{align*}
   \tilde{\phi}& = \phi\mp\/u \\
   \tilde{\r}\,\cos\,(\tilde{\th}\mp\/\tilde{\phi}) & 
         = \r\,\cos\,(\th \mp\/\phi+u) \pm\/ x_2  \\
   \tilde{\r}\,\sin\,(\tilde{\th}\mp\/\tilde{\phi}) & 
         = \r\,\sin\,(\th \mp\/\phi+u) \mp\/ x_1\,.
\end{align*}
From eqs.~(\ref{-inner}) and~(\ref{iso-aut}) it follows that the rank of
$\textnormal{Ad}_{g^{-1}}F-1$ is
\begin{equation*}
    \textnormal{rank}\,\big( \textnormal{Ad}_{g^{-1}}R^\Om_++1\big) = \left\{
       \begin{array}{lll}
      4 & \textnormal{if} &  \tilde{\phi}\neq\/(2n+1)\pi \\[9pt]
      2 & \textnormal{if} &  \tilde{\phi}=(2n+1)\pi
      \end{array} \right.
\end{equation*}
for $F=\!-R^\Om_+$  and
\begin{equation*}
   \textnormal{rank}\,\big( \textnormal{Ad}_{g^{-1}}R^\Om_-+1\big) = \left\{
     \begin{array}{lll}  
        3 & \textnormal{if}   
        & \tilde{\r} \sin\!\big(\tilde{\th} +\frac{\tilde{\phi}}{2}\big)
                                                       \!\neq 0 \\[9pt]
        1 & \textnormal{if} 
               &  \tilde{\r} \sin\!
               \big(\tilde{\th} +\frac{\tilde{\phi}}{2}\big)=0\,
      \end{array} \right.
\end{equation*}
for $F=\!-R^\Om_-$. We see that in both cases the rank of\,
$\textnormal{Ad}_{g^{-1}}F-1$ \,is not constant over ~$G_{\textnormal{\sc
    nw}}$.  The fields\, $t_a=(\textnormal{Ad}_{g^{-1}}F-1)T_a$ \,hence do not
provide a distribution on~$G_{\textnormal{\sc nw}}$.

\section*{Acknowledgment}

The authors are grateful to MEC and UCM-BSCH, Spain for partial support
through grants FPA2008-04906 and 910770(GR58/08). RHR acknowledges the Ram\'on
y Cajal Program, Spain for support.

\end{document}